\documentclass{svproc}
\usepackage{url}
\usepackage{graphicx}
\usepackage{xcolor}
\usepackage{tabularx}
\usepackage{cite}

\usepackage[autostyle]{csquotes}
\MakeOuterQuote{"}
\usepackage{booktabs}
\usepackage{authblk}

\begin{document}
\mainmatter              
\title{Can you always reap what you sow? Network and functional data analysis of VC investments in health-tech companies}
\titlerunning{Can you always reap what you sow?}  
%
\author{Christian Esposito\inst{1,2,*} \and Marco Gortan\inst{3,*} \and Lorenzo Testa\inst{1,2,*} \and Francesca Chiaromonte\inst{1,4} \and Giorgio Fagiolo\inst{1} \and Andrea Mina\inst{1,5} \and Giulio Rossetti\inst{6}}
\authorrunning{Esposito, Gortan, Testa et al.} 
\institute{Inst. of Economics \& EMbeDS, Sant'Anna School of Advanced Studies, Pisa, Italy
\and
Dept. of Computer Science, University of Pisa, Pisa, Italy
\and
Dept. of Finance, Bocconi University, Milan, Italy
\and 
Dept. of Statistics \& Huck Institutes of the Life Sciences, Penn State University, University Park, USA
\and
Centre for Business Research, University of Cambridge, Cambridge, UK 
\and
KDD Lab. ISTI-CNR, Pisa, Italy
\\ $*$ These authors contributed equally
}

\maketitle              

\begin{abstract}
"Success" of firms in venture capital markets is hard to define, and its determinants are still poorly understood. We build a bipartite network of investors and firms in the healthcare sector, describing its structure and its communities. Then, we characterize "success" introducing progressively more refined definitions, and we find a positive association between such definitions and the centrality of a company. In particular, we are able to cluster funding trajectories of firms into two groups capturing different "success" regimes and to link the probability of belonging to one or the other to their network features (in particular their centrality and the one of their investors). We further investigate this positive association by introducing scalar as well as functional "success" outcomes, confirming our findings and their robustness.

\keywords{network analysis, functional data analysis, success analysis, venture capital investments}
\end{abstract}

\section{Introduction}
Many phenomena may be described through networks, including investment interactions between bidders and firms in venture capital (VC) markets \cite{liang2016} and professional relationships among firms \cite{bonaventura2020}.
Risk capital is an essential resource for the formation and growth of entrepreneurial venture and venture capital firms are often linked together in a network by their joint investments in portfolio companies \cite{bygrave1988}. Through connections in such a network, they exchange resources and investment opportunities with one another. Many studies show the impact of network dynamics on investments, raising efficiency \cite{wetzel1987informal} and providing precious information when there is a great level of information asymmetry \cite{fiet1995reliance}. Also, differentiating connection types and avoiding tight cliques appear to help the success of an investor by providing more diverse information and reducing confirmation bias \cite{bygrave1988}.

CB Insights \cite{cbins} provides records of all transactions in venture capital markets from 1948. Since data until 2000 are partial and discontinuous, we focus on the period 2000-2020, in order to minimize the impact of missing data on our analysis. Additionally, since different sectors may be characterized by different investment dynamics \cite{dushnitsky2006does}, we focus on the healthcare sector, which is of great importance and has shown to be less sensitive to market oscillations \cite{pisano2006}. This stability is also shared by returns of life science VC, where investments have a lower failure rate but are at the same time less likely to generate "black-swan" returns \cite{booth2011defense}, offering more consistency but a lower likelihood of achieving billion-dollars evaluations. 

While the number of exits through an IPO or through a trade sale can be seen as a proxy for the success of an investor \cite{hege2003determinants}, there are instead different definitions of "success" for startups, but a common factor seems to be the growth rate of the company \cite{santi2017lit}. Our work aims to understand whether network features may affect "success" of investments in healthcare firms. In order to investigate this, we introduce progressively more nuanced definitions of "success", and analyze them with increasingly sophisticated statistical tools. 

The paper is organized as follows. Section \ref{sec:char} introduces and characterizes a network of investors and firms, describing its structure and salient properties, including the communities emerging from its topology. Then, Section \ref{sec:success} focuses on the definition and analysis of "successful" firms. We first characterize "success" by looking at the funding trajectories of each firm, clustering these trajectories into two broad groups capturing a high and a low funding regime. The binary cluster membership labels provide a first, rough definition of "success". We run a logistic regression in order to explain "success" defined in this fashion with statistics computed on the network itself. We then move to more complex characterizations of "success": the total amount of money raised (a scalar) and the funding trajectory itself (a functional outcome). We run regressions also on these outcomes, to validate and refine our previous results. Finally, we discuss main findings and provide some concluding remarks in Section \ref{sec:disc}.

\section{Network characterization}
\label{sec:char}

The 83258 agents in the healthcare sector are divided into two broad categories: 32796 bidders, or investors, and 50462 firms. Companies open investment calls in order to collect funds; investors answer such calls and finance firms. Each deal, i.e. each transaction from an investor to a company, is recorded in the CB Insights' database. This market dynamics can be described by a \textit{bipartite network}, which indeed is built on the notion of dichotomous heterogeneity among its nodes. In our case, each node may be a firm or an investor, respectively. An undirected link exists between two nodes of different kinds when a bidder has invested into a firm. Of course, given the possibility for an investor to finance the same firm twice, the bipartite network is also a \textit{multi-graph}. By knowing the date in which investments are made, we can produce yearly snapshots of the bipartite network. A company (investor) is included in a snapshot of a certain year only when it receives (makes) an investment that year. By projecting the bipartite network onto investors and firms, we produce the two projected graphs which are used to compute all the node statistics described in Table \ref{tab:variables}. As the bipartite network is a multi-graph, defining projections on a subset of nodes requires an additional assumption. Specifically, we project the bipartite graph onto firms by linking them in a cumulative fashion: we iteratively add to each yearly projected snapshot a link between two companies in which a bidder has invested during that year. Concerning the projection of the bipartite network onto investors, we link two bidders whenever they invest in the same company in the same financing round. 

\begin{table}[t]
    \centering
    \caption{Statistics computed on the projected graphs of investors and firms. Before running regressions in Section~\ref{sec:success}, left-skewed variables are normalized through log-transformation.}
    \label{tab:variables}
    \begin{tabularx}{\textwidth}{XX}
    \toprule
    \textbf{Variable} & \textbf{Network meaning}\\
    \midrule
    Degree centrality & Influence \\
    Betweenness centrality \cite{hannan1977population} & Role within flow of information \\
    Eigenvector centrality \cite{bonacich1987power} & Influence \\
    VoteRank  \cite{zhang2016identifying} & Best spreading ability \\
    PageRank \cite{page1999pagerank}& Influence \\
    Closeness centrality \cite{freeman1978centrality} & Spreading power (short average distance from all other nodes) \\
    Subgraph centrality \cite{estrada2005subgraph} & Participation in subgraphs across the network \\
    Average neighbor degree \cite{barrat2004architecture} & Affinity between neighbor nodes \\
    Current flow betweenness centrality \cite{newman2005measure} & Role within flow of information\\
    \bottomrule
    \end{tabularx}
\end{table}

Roughly 75\% of the companies in the network projected onto firms are North American and European (around 55\% belong to the US market), while the remaining 25\% is mostly composed of Asian companies. Around 60\% of the companies operate within the sub-sectors of medical devices, medical facilities and biotechnology -- the pharmaceutical sub-sector alone accounts for 20\% of the network. As of August, 2021, roughly of 80\% the companies in the network are either active or acquired, with the remaining portion being inactive or having completed an IPO. We witness turnover of the active companies through the years, but this is expected: a company's status is evaluated as of 2021, and it is more likely to observe a dead company among those that received investments in 1999 than in 2018. Indeed, both death and IPO represent the final stage of the evolution of a company, so those that received funding in earlier years are more likely to have already reached their final stage. Finally, we do not observe marked changes in terms of graph sub-sectoral composition: the relative share of each sub-sector is rather stable through the years, with the exception of an increase in the shares of the internet software and mobile software sub-sectors (from 1\% in 1999 to 8\% in 2019 and from 0\% in 1999 to 5\% in 2019, respectively). 

\subsection{Communities}

By employing the Louvain method \cite{blondel2008fast}, we identify meso-scale structures for each yearly snapshot of the network projected onto firms. For each year, we rank communities by their size, from the largest to singletons. We then compare the largest communities across years, by looking at their relative sub-sectors, status and geographical composition.

While the specific nodes in the biggest communities may vary throughout the years, we notice a relative stability in their features. The largest communities (which contain between 13\% and 20\% of the nodes) reflect the status composition of the general network, downplaying unsuccessful companies and giving higher relative weight to IPO ones, showing just a variation between acquired and active companies across years (i.e.~active companies are relatively over-represented in more recent largest communities than in older ones). Considering geographical information, the largest communities comprise mainly US companies, with an under-representation of other continents. This trait is quite consistent through the years, with the exception of 
two years (2013-2014). With respect to sub-sectors, the largest communities mainly contain medical device and biotechnology companies, and they are quite consistent through the years in terms of sub-sectoral composition. 

The second largest communities (containing between 10\% and 14\% of nodes in the network) have a less consistent sub-sectoral composition through the years, although it is worth highlighting that they comprise companies operating within software and technology. Geographically, we are still witnessing communities of mostly US-based companies, although 5 years out of 20 show a remarkable (roughly 80\%) presence of European companies. Finally, status composition is balanced between active and acquired until the later years, when active companies predominate within the second largest communities. IPOs are not present, while there are, in a small percentage (between 5\% and 20\%), dead startups.

Finally, the third largest communities (containing between 7\% and 12\% of the nodes) present a clear change within the period considered: in the first ten years, they mostly comprise failed or acquired European companies within the fields of biotechnology and drug development, while, in the second decade, they comprise active US companies within the fields of medical devices and medical facilities. 

\section{Success analysis}
\label{sec:success}
Given the bipartite network and its projections, we now turn to the analysis of success and of its main drivers. Because of the elusiveness of the definition of "success", we proceed in stages -- considering progressively more refined outcomes and comparing our findings. Moreover, since many of the records available in the CB Insights' data set are incomplete, and our aim is to capture the temporal dynamics leading a firm to succeed, we further restrict attention to those companies for which full information is available on birth year, healthcare market sub-sector and investment history for the first 10 years from founding. Although this filtering may introduce some biases, it still leaves us with a sizeable set of 3663 firms belonging to 22 different sub-sectors.

Notably, we restrict our focus also in terms of potential predictors, due to the fact that our collection of network features exhibits strong multicollinearities. By building a feature dendrogram (Pearson correlation distance, complete linkage) and by evaluating the correlation matrix, we reduce the initial set to four representatives. In particular, we select two features related to the investors' projection (the maximum among the degree centralities of the investors in a company and the maximum among their current flow betweenness centralities, both computed in the company's birth year) and two features computed on the firms' projection (a company's eigenvector and closeness centralities, computed in the year in which the company received its first funding). 

\begin{figure}[t]
    \centering
    \includegraphics[width=0.8\textwidth]{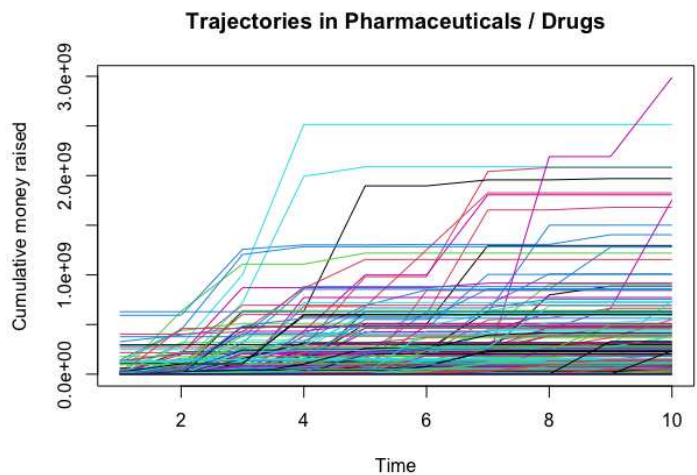}
    \caption{Money raised cumulatively as a function of time, shown for 319 firms in the pharmaceuticals and drugs sub-sector. Funding trajectories are constructed over a period of $10$ years since birth, and aligned using birth years as registration landmarks.}
    \label{fig:aligned_pharma}
\end{figure}

Each firm has its own funding history: after its birth, it collects funds over the years, building a \textit{trajectory} of the amount of money it is able to attract. We treat these trajectories as a specific kind of structured data, by exploiting tools from a field of statistics called \textit{Functional Data Analysis} (FDA) \cite{ramsey2005functional}, which studies observations that come in the form of functions taking shape over a continuous domain. In particular, we focus on the \textit{cumulative} function of the money raised over time by each company. As an example, Figure \ref{fig:aligned_pharma} shows 319 such cumulative functions, for the firms belonging to the pharmaceuticals and drugs sub-sector. Trajectories are \textit{aligned}, so that their domain ("time") starts at each company's birth (regardless of the calendar year it corresponds to). By construction, these functions exhibit two characterizing properties: first, they are monotonically non-decreasing; second, they are step functions, with jumps indicating investment events. 

\begin{figure}[t]
    \centering
    \includegraphics[width=0.9\textwidth]{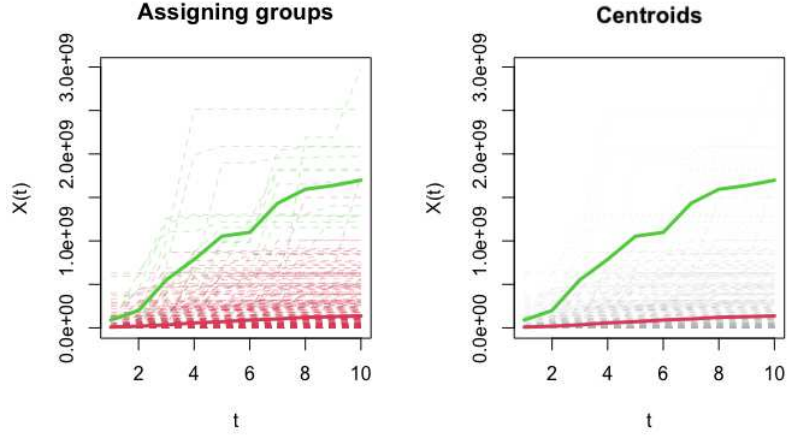}
    \caption{$k$-means clustering ($k=2$) of the funding trajectories of firms belonging to the pharmaceuticals and drugs sub-sector. The green and red dashed lines represent firms in the high ("successful") and low regimes, respectively. Bold curves represent cluster centroids. To aid their visualization, centroids are shown again in the right panel with individual trajectories in gray.}
    \label{fig:clustered_pharma}
\end{figure}

Our first definition of success is based on separating these trajectories into two regimes characterized by high (successful) vs. low investment patterns: the first runs at high levels, indicating successful patterns, and the second at low levels. Because of heterogeneity among healthcare sub-sectors, we accomplish this by running a \textit{functional k-means clustering} algorithm \cite{jacques2014func, hartigan1979} with $k=2$, separately on firms belonging to each sub-sector. 
As an example, companies belonging to the sub-sector of pharmaceuticals and drugs are clustered in Figure \ref{fig:clustered_pharma}. Throughout all sub-sectors, the algorithm clusters $89$ firms in the high-regime group and $3574$ in the low-regime one. 

This binary definition of "success" turns out to be rather conservative; very few firms are labeled as belonging to the high investment regime. Consider the logistic regression
%
\begin{equation}
\label{eq:log}
    \log\left(\frac{P(y_i=1)}{1-P(y_i=1)}\right)=\beta_0+\sum_{j=1}^{p} \beta_j x_{ij} \quad i=1,\dots n
\end{equation}
where $n$ is the number of observations, $y_i$, $i=1,\dots n$, are the binary responses indicating membership to the high ($y_i=1$) or low ($y_i=0$) regime clusters; $\beta_0$ is an intercept and $x_{ij}$, $i=1,\dots n$ and $j=1,\dots,p$ ($p=4$), are the previously selected scalar covariates.

\begin{figure}[t]
    \centering
    \includegraphics[width=\textwidth]{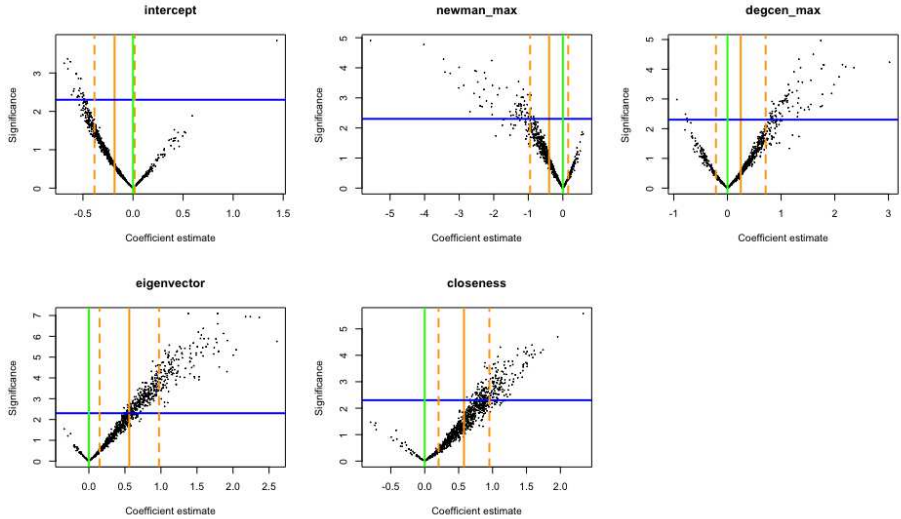}
    \caption{Scatter plots of logistic regression coefficient estimates (horizontal) and significance (vertical; $-log(p$-$value)$). Each point represents one of $1000$ fits run on data balanced by subsampling the most abundant class. Orange solid line mark averages across the fits, and orange dashed lines $\pm1$ standard deviations about them. Green solid lines mark $0$ on horizontal axes. Blue line mark significance values associated to a p-value of $0.1$.}
    \label{fig:est_coef}
\end{figure}

If we fit this regression on our unbalanced data, results are bound to be unsatisfactory and driven by the most abundant class. Running such a fit, one obtains an explained deviance of only $0.10$. To mitigate the effects of unbalanced data \cite{haibo2009}, we randomly subsample the most abundant class (the low-regime firms) as to enforce balance between the two classes, and then run the logistic regression in Equation \ref{eq:log}. We repeat this procedure 1000 times, recording estimated coefficients, associated p-values and explained deviances. The average of the latter across the 1000 replications is substantially higher than on the unbalanced fit, reaching $0.18$ 
(some fits produce deviance explained as high as $0.45$). Moreover, we can investigate significance and stability of the coefficient estimates through their distribution across the repetitions. Figure \ref{fig:est_coef} shows scatter-plots of these quantities, suggesting that the two variables related to the firms' centrality have a modest yet stable, positive impact on the probability of belonging to the high-regime cluster. This is not the case for the variables related to the investors' centrality.

This first evidence of a positive relationship between the success of a firm and its centrality, or importance (in a network sense) is promising. However, the binary definition of "success" we employed is very rough -- and the unbalance in the data forced us to run the analysis relying on reduced sample sizes ($89+89=178$ observations in each repeated run). Thus, we next consider a scalar proxy for "success", which may provide a different and potentially richer perspective. Specifically, we consider the cumulative end point of a firm's funding trajectory, i.e. the total value of the investment received through its temporal domain.

For this scalar response, we run a \textit{best subset selection} \cite{friedman2005elements} considering all the network features in our initial set -- not just the $4$ selected to mitigate multicollinearity prior to the logistic regression exercise. Notably, despite the substantial change in the definition of "success", results are in line with those from the logistic regression. Indeed, the first selected variable, when the predictor subset is forced to contain only one feature, is the eigenvector centrality of firms. When the predictor subset size is allowed to reach $4$, the features selected are the closeness and the VoteRank of the firm, and the maximum current flow betweenness centrality among its investors (computed on the firm's birth year). Thus, the only difference compared to our previous choice is the selection of the firms' VoteRank centrality instead of the maximum among the investors' degree centrality. We compare the two alternative selections of four features as predictors of the scalar "success" response fitting two linear models of the form:
\begin{equation}
    y_i =\beta_0+\sum_{j=1}^{p} \beta_j x_{ij} +\epsilon_i \quad i=1,\dots n
\end{equation}
where $n$ is the number of observations, $y_i$, $i=1,\dots n$, are the scalar responses (aggregate amount of money raised); $\beta_0$ is an intercept; $x_{ij}$, $i=1,\dots n$ and $j=1,\dots,p$ ($p=4$), are the scalar covariates belonging to one or the other subset and $\epsilon_i$, $i=1,\dots n$, are i.i.d. Gaussian model errors. As shown in Table \ref{tab:linreg}, the maximum degree centrality among a firm's investors is not statistically significant. 
Surprisingly, the maximum among investors' current flow betweenness centralities is significantly negative,  but its magnitude is close to 0. In contrast, the firms' closeness and eigenvector centralities are positive, statistically significant and sizeable. This is in line with what we expected, since it is reasonable to think that knowledge may indirectly flow from other startups through common investors, increasing the expected aggregate money raised. Finally, the firms' VoteRank centrality appears to have a negative, statistically significant impact on the aggregate money raised. This should not be surprising, given that the higher the VoteRank centrality is, the less influential the node will be. The variance explained by the two models is similar and still relatively low ($R^2 \approx 0.13)$, which may be simply due to the fact that network characteristics are only one among the many factors involved in a firm's success \cite{dosilimmancabile1994}. Nevertheless, the results obtained here through the scalar "success" outcome are consistent with those obtained through the binary one and logistic regression.

\begin{table}[t]
\centering 
\caption{Linear regressions of aggregate money raised on 
two sets of predictors. All variables are scaled and some are log-transformed (as indicated parenthetically).} 
\label{tab:linreg} 
\begin{tabular}{@{\extracolsep{5pt}}lcc} 
\toprule
 & \multicolumn{2}{c}{\textit{Dependent variable:}} \\ 
\cline{2-3} 
\\[-1.8ex] & \multicolumn{2}{c}{Aggregate money raised (log)} \\ 
\\[-1.8ex] & (1) & (2)\\ 
\midrule
 newman\_max & $-$0.065$^{**}$ & $-$0.072$^{*}$ \\ 
  & (0.030) & (0.041) \\ 
  & & \\ 
 voterank (log) & $-$0.140$^{***}$ &  \\ 
  & (0.033) &  \\ 
  & & \\ 
 degcen\_max (log) &  & 0.050 \\ 
  &  & (0.040) \\ 
  & & \\ 
 closeness & 0.126$^{***}$ & 0.130$^{***}$ \\ 
  & (0.037) & (0.030) \\ 
  & & \\ 
 eigenvector (log) & 0.214$^{***}$ & 0.255$^{***}$ \\ 
  & (0.034) & (0.028) \\ 
  & & \\ 
 Constant & 0.113$^{***}$ & 0.062$^{**}$ \\ 
  & (0.030) & (0.025) \\ 
  & & \\ 
\midrule \\[-1.8ex] 
Observations & 1,118 & 1,364 \\ 
R$^{2}$ & 0.136 & 0.127 \\ 
Adjusted R$^{2}$ & 0.133 & 0.125 \\ 
Residual Std. Error & 0.992 (df = 1113) & 0.923 (df = 1359) \\ 
F Statistic & 43.951$^{***}$ (df = 4; 1113) & 49.458$^{***}$ (df = 4; 1359) \\ 
\bottomrule
\textit{Note:}  & \multicolumn{2}{r}{$^{*}$p$<$0.1; $^{**}$p$<$0.05; $^{***}$p$<$0.01} \\ 
\end{tabular} 
\end{table} 

Our scalar outcome (aggregate money raised) has its own drawbacks. In particular, it implicitly assumes that the right time to evaluate success and investigate its dependence on network features is, cumulatively, at the end of the period considered (10 years). Note that this translates into a 10-year gap between the measurement of network features and financial success. 

\begin{figure}[t]
    \centering
    \includegraphics[width=\textwidth]{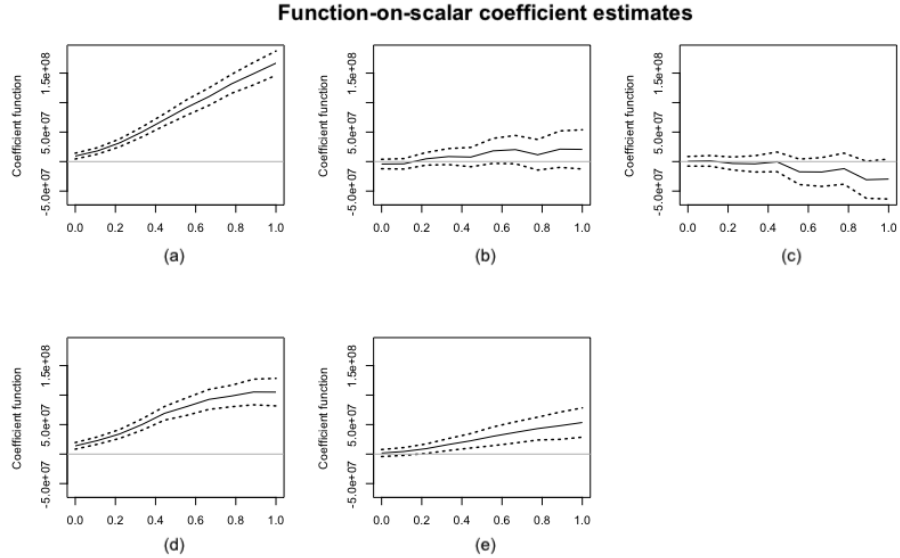}
    \caption{Function-on-scalar regression, coefficient curve estimates. (a) intercept function (this can be interpreted as the sheer effect of time on the response); (b) maximum degree centrality among investors (company's birth year); (c) maximum across investors' current flow betweenness centrality (company's birth year); (d) company's eigenvector centrality; (e) company's closeness centrality. Dotted lines represent confidence bands. All the covariates are standardized.}
    \label{fig:fos_coef}
\end{figure}

Although this issue could be approached relying on additional economic assumptions, we tackle it refining the target outcome and considering the full funding trajectories -- instead of just their end point. This requires the use of a more sophisticated regression framework from FDA; that is, function-on-scalar regression \cite{kokoszka2017introduction}. In particular, we regress the funding trajectories on the same two sets of covariates considered in the scalar case above. The equation used for function-on-scalar regression is:

\begin{equation}
    \label{eq:fun_reg}
        Y_i(t) = \beta_0(t) + \sum_{j=1}^{p} \beta_j(t) x_{ij} + \epsilon_i(t) \quad i=1,\dots n
    \end{equation}
where $n$ is the number of observations; $Y_i(t)$, $i=1,\dots n$, are the aligned funding trajectories; $\beta_0(t)$ is a functional intercept; $x_{ij}$, $i=1,\dots n$ and $j=1,\dots,p$ ($p=4$), are the scalar covariates belonging to the one or the other set, and $\epsilon_i(t)$, $i=1,\dots n$, are i.i.d. Gaussian model errors.

The regression coefficient of a scalar covariate in this model, $\beta_j(t)$, is itself a curve describing the time-varying relationship between the covariate and the functional response along its domain. Together with the functional coefficients, we also estimate their standard errors, which we use to build confidence bands around the estimated functional coefficients \cite{refund2016}. Coefficient curve estimates for the covariate set including the maximum investors' degree centrality are shown in Figure~\ref{fig:fos_coef} (results are very similar with the other set of covariates). The impacts of an increase in the maximum among the degree centralities and in the maximum among the current flow betweenness centralities of the investors in a firm are not statistically significant. Conversely, eigenvector and closeness centralities of firms have positive and significant impacts. The impact of the eigenvector centrality seems to be increasing during the first five years, reaching a "plateau" in the second half of the domain. These findings reinforce those obtained with the binary and scalar outcomes previously considered, confirming a role for firms' centrality in shaping their success.

\section{Discussion}
\label{sec:disc}
This paper exploits techniques from the fields of network and functional data analysis. We build a network of investors and firms in the healthcare sector and characterize its largest communities. Next, we progressively shape the concept of a firm's "success" using various definitions, and associate it to different network features. Our findings show a persistent positive relationship between the importance of a firm (measured by its centrality in the network) and various (binary, scalar and functional) definitions of "success". In particular, we cluster funding trajectories into a high ("successful") and a low regime, and find significant associations between the cluster memberships and firms' centrality measures. Then, we switch from this binary outcome to a scalar and then a functional one, which allow us to confirm and enrich the previous findings. Among centralities computed on the two network projections, our results suggest a preeminent role for those computed in the companies' projection. In particular, both a firm high closeness centrality, indicating a small shortest distances to other firms, and its eigenvector centrality, which may account for a firm's reputation, seem to be related to the propensity to concentrate capital.

Our analysis can be expanded in several ways. First, we limit our study to the healthcare sector, while it may be interesting to investigate other fields, or more healthcare firms based on the availability of more complete records. It would also be interesting to account for external data (e.g. country, sub-sector, etc.) in two ways. One the one hand, these information would be useful as to compute more informative statistics on the network topology. On the other hand, they may be used in our regression, to control for these factors.
Moreover, meso-scale communities may be analyzed in terms of their longitudinal evolution, as to characterize "successful" clusters of firms from a topological point of view. 

\section*{Acknowledgments}
F.C., C.E., G.F., A.M. and L.T. acknowledge support from the Sant'Anna School of Advanced Studies. 
F.C. acknowledges support from Penn State University. 
G.R. acknowledges support from the scheme "INFRAIA-01-2018-2019: Research and Innovation action", Grant Agreement n. 871042 "SoBigData++: European Integrated Infrastructure for Social Mining and Big Data Analytics".

%
%

\end{document}